\documentclass[twocolumn,showpacs,aps,prb,letterpaper,nofootinbib,raggedbottom,nobalancelastpage]{revtex4}
\usepackage{graphicx,bm}
\usepackage{amsmath}
\usepackage{color}

\begin{document}
\title{Highly efficient two photon generation  from a  coherently pumped quantum dot embedded in a microcavity}
\date{\today}
\author{J. K. Verma and P. K. Pathak}
\address{School of Basic Sciences,
Indian Institute of Technology Mandi, Kamand, H.P. 175005, India}
\begin{abstract}
We propose a scheme to realize a highly efficient solid state source of photon pairs using cavity-assisted stimulated Raman adiabatic passage
(STIRAP) in a single quantum dot, where a single photon from pump pulse and two stokes photons from cavity mode drives the Raman transition. The Autler-Townes
doublet, generated by using a resonant continuous wave laser between biexciton and exciton states, and two-photon-resonant transition through strongly coupled cavity mode are
utilized to facilitate (1+2)type Raman transition in the quantum dot. We show in the case of weak pump although the probability of generating two photons in cavity mode is small without cavity damping but two-photon-resonant emission is enhanced by cavity damping within strong coupling regime.
We also discuss spectrum of the generated photon pair and photon-photon correlations in the generated photon pair. The efficiency of two photon source could be more than 80\% in current experimental conditions.
\end{abstract}
\maketitle
\section{Introduction}
Two photon sources had wide range of applications in atomic physics \cite{atomic}, quantum metrology\cite{metrology}, quantum information processing\cite{qi},
quantum cryptography\cite{crypto}, in generating entangle photons\cite{entpair} and heralded single photon sources\cite{hsps}.
Until now, photon pairs employed in most of these experiments are generated by a parametric down conversion (PDC)\cite{PDC} or by four-wave mixing in
a nonlinear optical crystal\cite{rmp}. However, the number of generated photon pairs by such sources is random and the generation or collection efficiency is very low.
Recently there have been a lot of interest in generating source of precise single photon pair\cite{photopair,recent,laussy} on demand, particularly
for scalable quantum information processing and photonic technologies. Therefore sources using single emitters has become of great importance.
The efficiency of these sources is greatly enhanced by strong coupling with a high quality cavity, as a result two photon lasers\cite{twophlaser}
and masers\cite{twophmazer} have been realized. For developing "on chip" scalable solid-state photon sources, quantum dots (QDs) have been emerged
as a potential candidate. The QD can be strongly coupled in a photonic microcavity and can be grown precisely at the desired spatial location\cite{cqed}.

Two photon emission is a weak nonlinear process and is only significant when single photon processes
are suppressed in the emitters. In a single emitter of
upper energy level $|u\rangle$ and ground energy level $|g\rangle$, the resonant two-photon emission occurs at $\omega_{ug}=2\omega$; where $\omega_{ug}$ is the
transition frequency and $\omega$ is frequency of the emitted photons. The process proceeds through an intermediate state $|i\rangle$ which is far off-resonant to
the single photon transition. In quantum dots (QDs) this situation occurs naturally because of large biexciton binding energy, when a single mode cavity having
frequency half of the biexciton to ground state transition frequency is coupled\cite{recent,laussy}. So far, the two-photon emission observed from a single QD coupled
with a photonic crystal cavity rely on incoherent pumping\cite{recent}. The properties of the emitted photon pair strongly depend on pumping mechanism and in an
incoherently pumped source efficiency and purity of emitted photons are degraded\cite{incoherent}. Such limitations of incoherent pumping have also been noticed by Ota et. al.\cite{recent} in the
demonstration of two photon emission using single QD coupled with photonic crystal cavity. Because of inevitable excess scattering from pump laser,
coherent excitation of exciton levels in QD-cavity systems has been challenging task and it becomes very hard to differentiate between scattered photons and the emitted photons.
There have been some remarkable experiments demonstrating methods for coherent pumping of single QD using orthogonal excitation detection method
in the photonic planar micro-cavity\cite{flagg} and micropillar cavity\cite{ates}. In this paper, we propose a scheme
for generating source of single photon pairs using coherent pumping scheme which opens the possibility of stimulated (1+2)type Raman adiabatic passage (STIRAP) in
a single QD embedded in
a semiconductor photonic cavity\cite{pathak,ramandot1,ramandot2}. Recently, STIRAP has been implemented using negatively charged QD for generating single photons with 99.5\% indistinguishability\cite{ramandot2}. In our scheme the pump lasers not only have a large difference in frequency but also orthogonally polarized than the emitted photons.
\section{Model for the (1+2) type stimulated Raman adiabatic passage in QD-cavity system}
\label{Sec:Theory}
\begin{figure}[t!]
\centering
\includegraphics[height=2.2in]{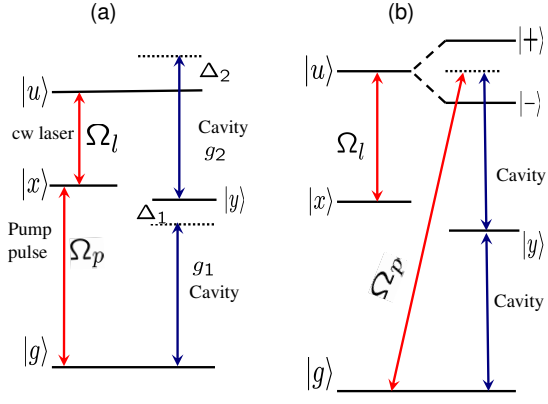}
\vspace{-0.1cm}
\caption{(Color online)
(a) Schematic of the photon pair source using the Raman adiabatic passage in a QD, using a CW laser ($\Omega_l$) and a pump pulse laser($\Omega_p(t)$),
and (b) the equivalent scheme in the dressed state picture.
 } \label{fig1}
\end{figure}
We consider a single quantum dot embedded in a single mode photonic microcavity. The quantum dot has two exciton energy levels
$|x\rangle$, $|y\rangle$, a biexciton energy level $|u\rangle$ and a ground state $|g\rangle$. In Fig.\ref{fig1}, we also consider the cavity mode,
having polarization along y-direction, is coupled through biexciton $|u\rangle$ to $|y\rangle$ transition and $|y\rangle$ to $|g\rangle$
transition with coupling constants $g_1$ and $g_2$ respectively. Because of large biexciton binding energy in QDs, resonant coupling to both transitions
$|u\rangle$ to $|y\rangle$ and $|y\rangle$ to $|g\rangle$ simultaneously with a single mode field is not possible.
Therefore we consider that the frequency of the cavity mode $\omega_c$ is chosen such that the two photon resonant condition
$\Delta_1+\Delta_2=0$, $|\Delta_1|>>g_1$, $|\Delta_2|>>g_2$ is satisfied. Here, $\Delta_1$ and $\Delta_2$ are the detunings of cavity mode to
the $|u\rangle$ to $|y\rangle$ transition and $|y\rangle$ to $|g\rangle$
transition respectively. We use notation $|\alpha,m\rangle$ for representing combined state of quantum dot and cavity,
where $\alpha$ represents the state of quantum dot and $m$ represent the number of photons in the cavity mode. Two photon transition between $|u,0\rangle$ to $|g,2\rangle$ via $|y,1\rangle$
takes place through cavity mode and the single photon transitions $|u,0\rangle\rightarrow|y,1\rangle$, $|y,1\rangle\rightarrow|g,2\rangle$,
and $|y,0\rangle\rightarrow|g,1\rangle$ have negligible probabilities. The quantum dot can also be pumped to biexciton state through less efficient two-photon
absorption from a pump laser. Furthermore, we propose an innovative approach for pumping mechanism, which makes two-photon emission in cavity mode
highly efficient, as follows. A resonant x-polarized cw-laser is applied between the energy states $|u\rangle$ and $|x\rangle$ with coupling constant
$\Omega_l$ and a x-polarized Gaussian pump pulse is applied between $|g\rangle$ and $|x\rangle$ with time dependent coupling $\Omega_p(t)$.
The hamiltonian for the system in the interaction picture is written as,
\begin{eqnarray}
&H  &= \hbar\Delta_p|x\rangle\langle x|+\frac{\hbar}{2}(\Delta_p+\Delta_l+\Delta_1-\Delta_2)|y\rangle\langle y|\nonumber\\
& \hbox{}&+ \ \hbar(\Delta_p+\Delta_l)|u\rangle\langle u|+\frac{\hbar}{2}(\Delta_p+\Delta_l-\Delta_1-\Delta_2)a^{\dag}a\nonumber\\
&\hbox{}&+\  \hbar\left[\Omega_l|u\rangle\langle x| +\Omega_p(t)|x\rangle\langle g|+g_1|y\rangle\langle g|a+g_2|u\rangle\langle y|a+H.c.\right],
\label{ham}
\end{eqnarray}
where $\Delta_p$ and $\Delta_l$ are the detunings of the pump pulse and the laser field and
 $H.c$ refers to Hermitian conjugate. Initially, the QD is
in the ground state $|g\rangle$ and the cavity mode has no photons.
The applied resonant cw-laser field between the $|x\rangle$ and $|u\rangle$, creates Autler-Townes doublets $|\pm\rangle=1/\sqrt{2}(|u\rangle\pm|x\rangle)$\cite{autler1,autler2}.
The energy separation between the states $|+\rangle$ and $|-\rangle$ is given by $\Omega_l$. The states $|\pm\rangle$ are equally coupled to ground state
$|g\rangle$ through dipole transition. Further, they are also coupled to state $|g,2\rangle$ through cavity induced two-photon resonant
transitions via state $|y,1\rangle$. The two-photon transitions $|+\rangle\rightarrow|y,1\rangle\rightarrow|g,2\rangle$ and
$|-\rangle\rightarrow|y,1\rangle\rightarrow|g,2\rangle$ interfere constructively and maximum population is transferred to $|g,2\rangle$ without populating
$|y,1\rangle$. The detuning for pump pulse becomes $\Delta_p\pm\Omega_l/2$ and the detuning for two-photon resonant transition becomes
$\Delta_1+\Delta_2\pm\Omega_l/2$ in dressed state picture. When a properly selected x-polarized gaussian pump pulse is applied between QD states $|g,\rangle$ and
$|x\rangle$, the population is faithfully transferred to $|g,2\rangle$ through (1+2)type Raman adiabatic passage\cite{stirap}, from where the photon pair is emitted from cavity mode.

For simulating the dynamics of the system, we perform Master equation calculations in the density matrix representation:
\begin{eqnarray}
\frac{\partial\rho}{\partial t}=-\frac{i}{\hbar}[H,\rho]-\frac{1}{2}\sum_{\mu}L_{\mu}^{\dag}L_{\mu}\rho -2L_{\mu}\rho L_{\mu}^{\dag}+\rho L_{\mu}^{\dag}L_{\mu} ,
\label{master}
\end{eqnarray}
where $L_{\mu}$ are the Lindblad operators, with
$L_1=\sqrt{\gamma_1}|x\rangle\langle u|$, $L_2=\sqrt{\gamma_1}|y\rangle\langle u|$, $L_3=\sqrt{\gamma_2}|g\rangle\langle x|$, and $L_4=\sqrt{\gamma_2}|g\rangle\langle y|$ correspond to the spontaneous decays, and $L_5=\sqrt{2\gamma_d}|u\rangle\langle u|$, $L_6=\sqrt{\gamma_d}|x\rangle\langle x|$, and $L_7=\sqrt{\gamma_d}|y\rangle\langle y|$ correspond to the dephasing of biexciton and exciton states. The emission of photons from the cavity mode is given by the Lindblad operator $L_8=\sqrt{\kappa}a$.
The optical Bloch equations using density operator(\ref{master}) and $\partial\langle i|\rho|j\rangle/\partial t=\dot{\rho}_{ij}$, are given by
\begin{eqnarray}
\label{dens1}
\dot{\rho}_{gg}=-i\Omega^*_p\rho_{xg}+i\Omega_p\rho_{gx}+\kappa\rho_{GG}+\gamma_2(\rho_{yy}+\rho_{xx}),\\
\dot{\rho}_{xx}=-i\Omega^*_l\rho_{ux}+i\Omega_l\rho_{xu}-i\Omega_p\rho_{gx}+i\Omega^*_p\rho_{xg}-\gamma_2\rho_{xx}\nonumber\\
+\gamma_1\rho_{uu},\\
\dot{\rho}_{uu}=-i\Omega_l\rho_{xu}+i\Omega^*_l\rho_{ux}-ig_2\rho_{Yu}+ig_2\rho_{uY}-2\gamma_1\rho_{uu}\\
\dot{\rho}_{YY}=-ig_2(\rho_{uY}-\rho_{Yu})+ig_1\sqrt{2}(\rho_{YG'}-\rho_{G'Y})\nonumber\\
-(\kappa+\gamma_2)\rho_{YY},\\
\dot{\rho}_{G'G'}=-ig_1\sqrt{2}(\rho_{YG'}-\rho_{G'Y})-2\kappa\rho_{G'G'},\\
\dot{\rho}_{GG}=-ig_1(\rho_{yG}-\rho_{Gy})+\gamma_2\rho_{YY}+2\kappa\rho_{G'G'}-\kappa\rho_{GG},\\
\dot{\rho}_{yy}=-ig_1(\rho_{Gy}-\rho_{yG})+\kappa\rho_{YY}+\gamma_1\rho_{uu}-\gamma_2\rho_{yy},\\
\dot{\rho}_{yG}=-i\Delta_1\rho_{yG}-ig_1(\rho_{GG}-\rho_{yy})-\frac{(\kappa+\gamma_2+\gamma_d)}{2}\rho_{yG}\nonumber\\
+\kappa\sqrt{2}\rho_{YG'},\\
\dot{\rho}_{xg}=-i\Delta_p\rho_{xg}-i\Omega^*_l\rho_{ug}-i\Omega_p(\rho_{gg}-\rho_{xx})\nonumber\\
-\frac{(\gamma_2+\gamma_d)}{2}\rho_{xg},\\
\dot{\rho}_{YG'}=-i\Delta_1\rho_{YG'}-ig_2\rho_{uG'}-ig_1\sqrt{2}(\rho_{G'G'}-\rho_{yy})\nonumber\\
-\frac{(3\kappa+\gamma_2+\gamma_d)}{2}\rho_{YG'},\\
\dot{\rho}_{uY}=-i\Delta_2\rho_{uY}-i\Omega_l\rho_{xY}-ig_2(\rho_{YY}-\rho_{uu})\nonumber\\
+ig_1\sqrt{2}\rho_{uG'}-\frac{(\kappa+2\gamma_1+\gamma_2+3\gamma_d)}{2}\rho_{uY},\\
\dot{\rho}_{ux}=-i\Delta_l\rho_{ux}-i\Omega_l(\rho_{xx}-\rho_{uu})+i\Omega^*_p\rho_{ug}-ig_2\rho_{Yx}\nonumber\\
-\frac{(2\gamma_1+\gamma_2+3\gamma_d)}{2}\rho_{ux},\\
\dot{\rho}_{ug}=-i(\Delta_p+\Delta_l)\rho_{ug}-i\Omega_l\rho_{xg}+i\Omega_p\rho_{ux}-ig_2\rho_{Yg}\nonumber\\
-(\gamma_1+\gamma_d)\rho_{ug},\\
\dot{\rho}_{uG'}=-i(\Delta_1+\Delta_2)\rho_{uG'}-i\Omega_l\rho_{xG'}-ig_2\rho_{YG'}\nonumber\\
+ig_1\sqrt{2}\rho_{uY}-(\kappa+\gamma_1+\gamma_d)\rho_{uG'},\\
\dot{\rho}_{Yx}=-i(\Delta_l-\Delta_2)\rho_{Yx}+i\Omega_l\rho_{Yu}+i\Omega^*_p\rho_{Yg}-ig_2\rho_{ux}\nonumber\\
-ig_1\sqrt{2}\rho_{G'x}-(\gamma_2+\gamma_d+\kappa/2)\rho_{Yx},\\
\dot{\rho}_{Yg}=-i(\Delta_p+\Delta_l-\Delta_2)\rho_{Yg}+i\Omega_p\rho_{Yx}-ig_2\rho_{ug}\nonumber\\
-ig_1\sqrt{2}\rho_{G'g}-\frac{(\gamma_2+\gamma_d+\kappa)}{2}\rho_{Yg},\\
\dot{\rho}_{xG'}=-i(\Delta_1+\Delta_2-\Delta_l)\rho_{xG'}-i\Omega^*_l\rho_{uG'}-i\Omega_p\rho_{gG'}\nonumber\\
+ig_1\sqrt{2}\rho_{xY}-\frac{(\gamma_2+\gamma_d+2\kappa)}{2}\rho_{xG'},\\
\dot{\rho}_{G'g}=-i(\Delta_p+\Delta_l-\Delta_1-\Delta_2)\rho_{G'g}+i\Omega_p\rho_{G'x}\nonumber\\
-ig_1\sqrt{2}\rho_{Yg}-\kappa\rho_{G'g},
\label{dens2}
\end{eqnarray}
where we have used notation $|g\rangle\equiv|g,0\rangle$, $|x\rangle\equiv|x,0\rangle$, $|y\rangle\equiv|y,0\rangle$, $|u\rangle\equiv|u,0\rangle$, $|Y\rangle\equiv|y,1\rangle$, $|G\rangle\equiv|g,1\rangle$ and $|G'\rangle\equiv|g,2\rangle$. We solve the Eqs.(\ref{dens1})-(\ref{dens2}), for Gaussian pump pulse $\Omega_p(t)$ and calculate probabilities of different states  $\langle \alpha,m|\rho(t)|\alpha,m\rangle$ at any time $t$.

In our calculations we have used a fixed dephasing rate due to electron phonon coupling. Although, in QD-cavity systems electron phonon interactions have very complex nature and different mechanism such as pure dephasing\cite{tdeph1}, phonon-mediated transitions\cite{borri}, photon induced shake-up processes~\cite{shakeup} have been proposed to understand observed features in different experimental conditions. However, for coherently excited QDs coupling to continuum states in the wetting layer and shake-up processes due to charged excitons have been found negligible. Recent experiments at low temperatures, 5-10\,K, using low power coherent excitation in resonant QD-cavity systems, have found that phenomenological exponential dephasing and exponential radiative decay are very well fitted with the data of Rabi oscillations and spectral line widths. In a few remarkable experiments resonance fluorescence from doubly dressed QD\cite{ddressed} and the resonant coupling of Mollow sideband with cavity mode\cite{sideband} have also been observed. For coherently driven off-resonant QD-cavity systems excitation induced dephasing has also been found to play a significant role in Mollow triplet side bands broadening\cite{eid}. We predict that in our system excitation induced dephasing can not play significant role because through cavity mode only those two-photon resonant transitions which also satisfy Raman resonance condition with pump pulse can occur\cite{stirap}. We also remark that only the $y-$polarized transition satisfying the two-photon resonance condition from exciton states are possible, and thus the presence of other background states, e.g, charged excitons, do not affect the evolution of the system. In fact the two-photon Raman resonant transitions have been successfully implemented between two selected motional states of trapped ion within a manifold of motional states~\cite{motion}.

\section{Population dynamics and spectrum of the emitted photon pair}
\begin{figure}[h!]
\centering
\includegraphics[height=3in]{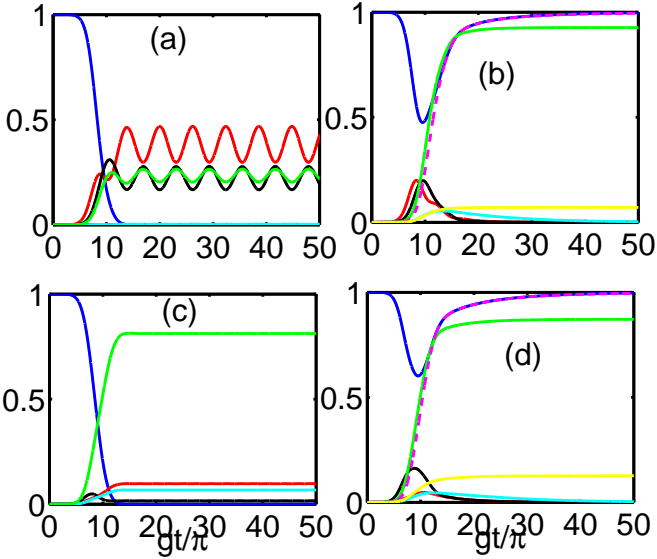}
\vspace{-0.1cm}
\caption{(Color online)
The population in $\langle g,0|\rho|g,0\rangle$(blue), $\langle x,0|\rho|x,0\rangle$ (red), $\langle u,0|\rho|u,0\rangle$ (black). In (a) and (c), the population in $\langle y,1|\rho|y,1\rangle$ (cyan), $\langle g,2|\rho|g,2\rangle$ (green). After introducing damping, $\gamma_1=\gamma_2=10^{-3}g$, $\gamma_d=10^{-2}g$, $\kappa=0.5g$, in (b) and (d), the population in $\langle y,0|\rho|y,0\rangle$ (cyan) and the probabilities $P_1=2\kappa\int_0^t\langle g,2|\rho(t')|g,2\rangle dt'$ (green), $P_2=\kappa\int_0^t\langle g,1|\rho(t')|g,1\rangle dt'$ (magenta dashed), $P_3=\kappa\int_0^t\langle y,1|\rho(t')|y,1\rangle dt'$ (yellow). The parameters are $g_1=g_2=g$, $\Delta_1=-\Delta_2=4g$, $\Omega_p(t)/g=A_p\exp[-(t-2.5\tau)/\tau^2]$, $g\tau/\pi=3.5$. For (a) \& (b) $\Omega_l/g=0.2$, $\Delta_p/g=0.14$, for (a) $A_p=0.13$, and for (b) $A_p=0.14$. For (c) \& (d) $\Omega_l/g=1$, $\Delta_p/g=0.455$, for (c) $A_p=0.28$, and for (d) $A_p=0.438$.
 } \label{fig2}
\end{figure}
Next, we present our results of numerical simulations in Fig.2. The y-polarized cavity mode satisfy two photon resonance $\Delta_1=-\Delta_2=4g$ and the dot is driven by x-polarized pump pulse and cw-laser. We consider two different regime, weak driving regime in which driving fields $\Omega_l,~\Omega_p(t)$ $\ll$ cavity damping $\kappa$, in Fig.2(b) and strong driving regime in which driving fields $\Omega_l,~\Omega_p(t)$ $>$ cavity damping $\kappa$ in Fig.2(d). In Figs. 2(a) and (c) results without spontaneous decay and cavity damping using same parameters as in (b) and (d), respectively, are plotted. In Figs. 2(a) and (c) required values of pump pulse amplitude is slightly lower than Figs. 2(b) and (d), respectively, for complete population transfer from $|g,0\rangle$ to $|g,2\rangle$. We consider generation of up to four photons in cavity mode ,i.e. up to state $|g,4\rangle$. However we found the probabilities of generating $|g,4\rangle$ state is  $<0.05$ in (a) and $<0.1$ in (c), which is effect of the photon blockade, as Raman resonance condition is not satisfied due to the stark shifts produced by the presence of the photons in cavity mode. By introducing finite cavity damping in (b) and (d), probability of generating four photons in cavity mode becomes negligible. Therefore, we have neglected coupling to such states in Figs 2(b) and (d) and considered up to two photon state $|g,2\rangle$. In Fig.2(a), when there is no damping ($\gamma_1=\gamma_2=\gamma_d=\kappa=0$) and the dot is weakly driven, the population from initial state $|g,0\rangle$ is completely transferred to states $|x,0\rangle$, $|u,0\rangle$ and $|g,2\rangle$ without populating any other state significantly. The population in $|g,2\rangle$ remains low ($\approx0.25$). However, when  the cavity damping such that $\kappa<g_1,g_2$ is introduced in Fig.2(b) the population in $|x,0\rangle$, $|u,0\rangle$ goes to zero with maximum around 0.2. We calculate probabilities of emitting photon $P_1=2\kappa\int_0^t\langle g,2|\rho(t')|g,2\rangle dt'$, $P_2=\kappa\int_0^t\langle g,1|\rho(t')|g,1\rangle dt'$, $P_3=\kappa\int_0^t\langle y,1|\rho(t')|y,1\rangle dt'$ from states $|g,2\rangle$, $|g,1\rangle$, and $|y,1\rangle$, respectively. We choose values of $\Omega_p(t)$ to achieve complete population transfer from initial state $|g,0\rangle$ in (a) and (c) and to achieve probability $P_2=1$ in (b) and (d). We find that $P_2\approx P_1+P_3$ in long time limit, which shows that the population in $|g,1\rangle$ is due to leakage of photon from $|g,2\rangle$ as well as from single photon transition $|y,0\rangle\rightarrow|g,1\rangle$. Therefore we can estimate the efficiency of two-photon transition $\eta=\lim_{t\rightarrow\infty}\frac{P_1(t)}{P_1(t)+P_3(t)}\approx\frac{P_1(t)}{P_2(t)}$. We find that in the weak driving limit even though the maximum population in $|g,2\rangle$ without damping is small but the efficiency of photon pair generation is very high ($\approx0.9$). In fact, without cavity damping population gets trapped in $|x,0\rangle$, $|u,0\rangle$ and $|g,2\rangle$ and in the presence of cavity damping population in $|g,2\rangle$ decays fast so more and more population gets transferred to $|g,2\rangle$. However the cavity-quantum dot system should be in strong coupling limit to achieve two photon resonant transitions. In the strong driving limit, the maximum population in $|g,2\rangle$ without damping is larger than 0.8 and similarly the efficiency $\eta\approx0.85$ for $\kappa=0.5g$.

\begin{figure}[h!]
\centering
\includegraphics[height=2.5in]{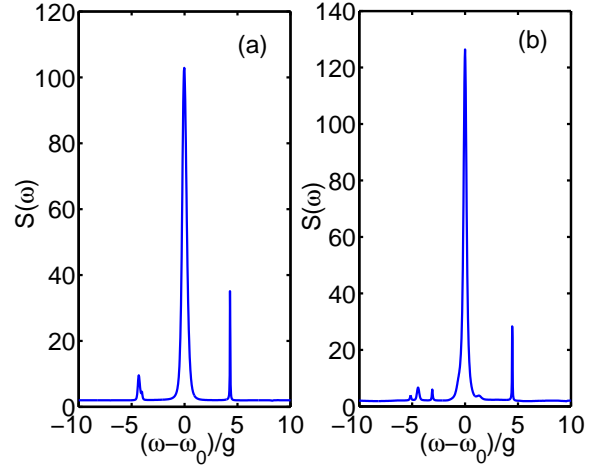}
\vspace{-0.1cm}
\caption{(Color online)
Spectrum of the emitted photons through cavity mode, (a) for parameters used in Fig.2(b) and (b) for parameters used in Fig.2(d).
 } \label{fig3}
\end{figure}
In Fig.3, we present calculated cavity mode spectrum $S(\omega)=\int_0^{\infty}dt\int_0^{\infty}d\tau a^{\dagger}(t)a(t+\tau)\exp{i\omega\tau}$. The two time correlation $a^{\dag}(t)a(t+\tau)$ is calculated using quantum regression theorem. We get one major peak at cavity frequency $\omega_c$ corresponding to two photon resonant emission and two tiny side peaks corresponding to single photon emission from cavity mode via cascaded decay of biexciton at $\omega-\omega_c=\Delta_1,\Delta_2$. Further in strong driving limit the peak corresponding to $\omega-\omega_c=\Delta_2$, splits into other peaks. The appearance of many peaks at $\omega-\omega_c=\Delta_2$ can we understood using dressed state picture\cite{ddressed}.

\section{photon-photon correlations}
\begin{figure}[h!]
\centering
\includegraphics[height=2.5in]{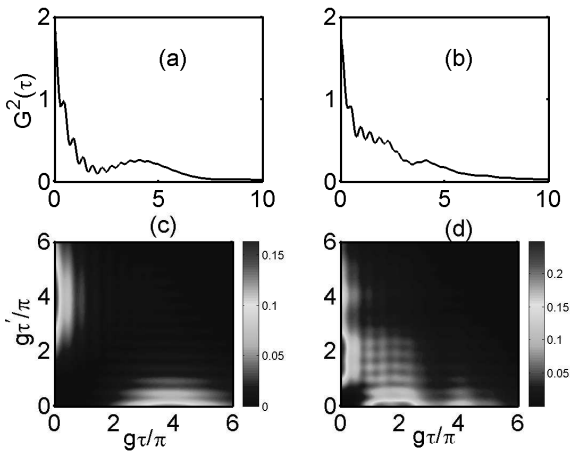}
\vspace{-0.1cm}
\caption{(Color online) Two photon correlation function $G^2(\tau)$ is plotted in (a), (b); and three photon correlation function $G^3(\tau,\tau^{\prime})$ is shown in (c), (d). The parameters for (a) and (c) are same
as in Fig.2(b) and parameters for (b) and (d) are same as in Fig.2(d).
 } \label{fig4}
\end{figure}
We present results of photon-photon correlations in Fig.4. We calculate two photon correlation $G^2(\tau)=\int_0^{\infty}dt a^{\dag}(t)a^{\dag}(t+\tau)a(t+\tau)a(t)$ and three photon correlation $G^3(\tau,\tau^{\prime})=\int_0^{\infty}dt a^{\dag}(t)a^{\dag}(t+\tau)a^{\dag}(t+\tau+\tau^{\prime})a(t+\tau+\tau^{\prime})a(t+\tau)a(t)$ using quantum regression theorem and Eq.(\ref{master}). In our system two types of processes are involved for photon emission, resonant two photon emission which leads to simultaneous two photon emission at cavity frequency and cavity modified spontaneous biexciton-exciton cascade decay leading to slow emission of two photons at $(\omega-\omega_c)\approx\Delta_1,\Delta_2$. In Fig.4 (a), (b), we find the first peak around $\tau\approx0$, corresponding to simultaneous resonant emission of two photons, followed by extended tail, corresponding to slow emission of two photons in cascaded decay. In Fig.4 (c), (d), corresponding three photon correlation function is plotted. Two main conclusion can be drawn from (c)\& (d)--fist there is no temporal overlapping between two photon resonant emission and cascaded decay, as $G^3$ has zero value for $\tau=\tau^{\prime}\approx0$.  In fact two photon resonant emission finishes first (see Fig.2(b)\& (d)). Second the nonzero values of $G^{3}$ lies along the axis, confirming that two photons in resonance process are emitted simultaneously and two-photons in cascaded decay are emitted slowly.

\section{Conclusions}
\label{Sec:Conclusions}
We have presented a new methods for efficient resonant excitation in a single neutral QD embedded in a photonic microcavity which enables (1+2)type STIRAP. The efficiency of generating photon pairs could be very large, thus open up the possibility of realizing a solid state "on demand" source of photon pair.
\section{Acknowledgements}
This work was supported by DST grant

\end{document}